\documentclass[preprint,12pt]{aastex}
\usepackage{epstopdf,natbib}

\newcommand{\name}{WISEPA J045853.90+643452.6}

\shorttitle{WISE Brown Dwarf Binaries}
\shortauthors{Gelino et al.}

\begin{document}

\title{WISE Brown Dwarf Binaries: The Discovery of a T5+T5 and a T8.5+T9 System\altaffilmark{1}}

\author{Christopher R.\ Gelino\altaffilmark{2},  J. Davy Kirkpatrick\altaffilmark{2}, Michael C. Cushing\altaffilmark{3}, 
Peter R. Eisenhardt\altaffilmark{3}, Roger L. Griffith\altaffilmark{2}, Amanda K. Mainzer\altaffilmark{3}, Kenneth A. Marsh\altaffilmark{2},
Michael F. Skrutskie\altaffilmark{4}, and Edward L. Wright\altaffilmark{5}}

\altaffiltext{1}{Some of the data presented herein were obtained at the
  W.M. Keck Observatory, which is operated as a scientific partnership
  among the California Institute of Technology, the University of
  California, and the National Aeronautics and Space Administration. The
  Observatory was made possible by the generous financial support of the
  W.M. Keck Foundation.}
\altaffiltext{2}{Infrared Processing and Analysis Center, California Institute of Technology, Pasadena, CA 91125, USA}
\altaffiltext{3}{NASA Jet Propulsion Laboratory, 4800 Oak Grove Drive, Pasadena, CA 91109, USA}
\altaffiltext{4}{Department of Astronomy, University of Virginia, Charlottesville, VA 22903, USA}
\altaffiltext{5}{Astronomy Department, University of California, Los Angeles, P.O. Box 951547, Los Angeles, CA 90095, USA}

\begin{abstract}
The multiplicity properties of brown dwarfs are critical empirical constraints for formation theories, while multiples themselves provide unique opportunities to test
evolutionary and atmospheric models and examine empirical trends. Studies using high-resolution imaging can not only uncover faint companions, but they can also be used to determine dynamical masses through long-term monitoring of binary systems.  We have begun a search for the coolest brown dwarfs using preliminary processing of data from the Wide-field Infrared Survey Explorer (WISE) and have confirmed many of the candidates as late-type T dwarfs. In order to search for companions to these objects, we are conducting observations using the Laser Guide Star Adaptive Optics system on Keck II. Here we present the first results of that search, including a T5 binary with nearly equal mass components and a faint companion to a T8.5 dwarf with an estimated spectral type of T9.
\end{abstract}


\section{Introduction}
Brown dwarfs are a natural link between extrasolar planets and low-mass stars. Their study in binary systems provides a method by which to obtain mass and radii measurements \citep{2001ApJ...560..390L, 2004ApJ...615..958Z, 2006Natur.440..311S, 2008ApJ...689..436L, 2009ApJ...692..729D} and a means to constrain atmosphere models regardless of the brown dwarf's age and metallicity \citep{2005ApJ...634..616L, 2007ApJ...657.1064M, 2010ApJ...710.1142B}.  While the transition between brown dwarfs and stars is well populated with objects  \citep{kirk2005}, there remains a temperature gap of several hundred K between the coldest brown dwarfs \citep{lucas2010,liu2011} and Jupiter.  

One method to find cool brown dwarfs is through deep imaging surveys.  Although the UKIRT Infrared Deep Sky Survey  \citep[UKIDSS;][]{warren2007,burningham2008,burningham2009,burningham2010,lucas2010} and Canada-France Brown Dwarfs Survey \citep[CFBDS;][]{delorme2008,delorme2010} surveys have already extended the T dwarf spectral sequence to cool brown dwarfs with spectral types T8 and later, the Wide-field Infrared Survey Explorer \citep[WISE;][]{wright2010} has pushed this to even cooler objects (Kirkpatrick et al., submitted; Cushing et al. in prep.).  The advantage of these surveys is that they cover large areas of the sky, thereby increasing the chance of discovering these hard-to-find objects.  The downside to these surveys is that they are generally designed to survey large areas of sky at the expense of deep images and, because the absolute magnitudes of brown dwarfs get fainter with decreasing effective temperatures, the effective temperatures of the coldest discoveries are limited by the depth of the survey.

A second way to find cool brown dwarfs is to search for them as binary companions to already cool, known brown dwarfs, such as those found in the wide-field surveys.  High resolution imaging surveys are the dominant means for finding companions and the most prolific, with $>$100 systems found to date \citep{burgasser2007}.  While they can probe to much fainter (and cooler) objects than wide area surveys, high resolution imaging programs are limited in the number of objects they can discover by the binary fraction of cool brown dwarfs and the limitations of ground-based adaptive optics systems.  

We have initiated a program to search for the coolest brown dwarfs using both of the above methods.  Specifically, we use newly discovered brown dwarfs from WISE \citep[][Kirkpatrick et al., submitted]{mainzer2011} and target them with high resolution imaging at Keck.  In Section~\ref{wise} we briefly summarize the WISE project and how it is sensitive to brown dwarfs.  Section~\ref{obs} details our Keck observing and data reduction procedures, and our results are given in Section~\ref{results}.

\section{Wide-field Infrared Survey Explorer\label{wise}}
The Wide-field Infrared Survey Explorer \citep[WISE;][]{wright2010} is a NASA Medium Class Explorer (MIDEX) mission that was designed to survey the entire sky simultaneously at four infrared wavelengths: 
3.4$\mu$m (W1), 4.6$\mu$m (W2), 12$\mu$m (W3), and 22$\mu$m (W4).   It was launched on 2009 December 14 from Vandenberg Air Force Base in California and, after a 1 month check-out period, it continuously obtained images of the sky. WISE completed a 4-band, first pass of the sky on 2010 July 17.  The telescope and arrays were kept cool by a two-stage cryostat; the cryogen in the outer cryostat was exhausted in early August 2010.   The increase in the thermal background caused the W4 array to saturate on  2010 August 8.   The W3 array became saturated on 2010 September 30, shortly after the coolant in the primary  cryostat was depleted.  Data collection for W1 and W2 ceased on 2011 February 1, after a second epoch of the sky was completed and briefly into the start of a third epoch.  

One of the primary goals of WISE is to identify ultracool brown dwarfs. In fact, the W1 and W2 bands were designed specifically to probe the deep CH$_4$ absorption band at $\sim$3.3$\mu$m and the region relatively free of opacity at $\sim$4.6$\mu$m in the spectra of cool brown dwarfs \citep{bsl2003}. Since the peak of the Planck function at these low temperatures is in the mid-IR, a large amount of flux emerges at 4.6$\mu$m, making the W1$-$W2 colors of cool brown dwarfs extremely red \citep[$>$2.0 mag; see Figure 12 in][]{wright2010}.  This very red W1$-$W2 color is almost unique among astronomical objects.

The WISE Point Source Working Database contains well over 1 billion entries.  We have spectroscopic follow-up observations for a small sample of our best brown dwarf candidates and have confirmed $\approx$80 of these objects as having spectral types cooler than T5, which more than triples the number of objects with spectral types T8 and cooler (Kirkpatrick et al., submitted).  In an effort to find brown dwarf binaries with even cooler, lower mass companions, we have observed with high-resolution imaging 9 of these spectroscopically confirmed, late-type brown dwarfs (T5-T9) to search for binary companions.

\section{Observations\label{obs}}
Our targets were selected from our list of spectroscopically confirmed WISE brown dwarfs \citep[][Kirkpatrick et al., submitted]{mainzer2011}.  We generally opted for brown dwarfs that were already very cool (spectral types T8 and later), though we did observe earlier types to fill holes during the night.  All of the WISE brown dwarfs are much too faint at optical wavelengths to serve as natural guide stars (NGS) for the Keck II Adaptive Optics (AO) system, so we required the use of the laser guide star (LGS) to provide the input for the wavefront corrector.  While the LGS can be pointed to virtually any position in the sky above 2 airmasses, the system needs a reference star to provide the tip-tilt corrections.  Tip-tilt reference stars can be as faint as $R\approx$18 mag for Keck II, which is still much brighter than our targets at this wavelength.  A search of the USNO-B catalog \citep{monet2003} showed that all but a few of our brown dwarfs had a suitably bright star within $\approx$60$\arcsec$, the limit for the Keck II AO system.  Table~\ref{tab_targs} presents the targets that were observed for this program.  The full WISE designations are given in the table in the form WISEPA Jhhmmss.ss$\pm$ddmmss.s and are truncated to WISE hhmm$\pm$ddmm hereafter.

High resolution imaging observations of our targets were obtained using the Keck II LGS-AO system \citep{wizinowich2006,vandam2006} with NIRC2 on the nights of 24 March 2010, 1 July 2010, and 27 December 2010 (UT).  Conditions were good during the March and July nights with seeing $<$1$\arcsec$; the December night was fraught with high humidity and instrument problems\footnote{The seeing on the December night was not measured because the seeing monitor was not running, nor did we take any non-AO images in our limited on-sky time.}.  The data were obtained by using a 3-point dither pattern that avoided the noisy, lower left quadrant of the array.  The pattern was repeated as needed, though with different dither offsets, to build up long exposures.    The Mauna Kea Observatories \citep[MKO;][]{simons2002,tokunaga2002} $H$ filter and narrow
plate scale (0.009942$\arcsec$/pixel for a single-frame field-of-view of $10\arcsec\times10\arcsec$) were used for the March and July observations because they provide the best compromise between image quality (i.e. Strehl ratio) and target-companion brightness contrast.  The wide camera (plate scale =  0.039686$\arcsec$/pixel) was used for the December night because the goal of that observation was to provide a deep exposure of the target with as many neighboring stars as possible in order to obtain good photometry.  Fortunately, even with this larger plate scale, the observations still provide a good measure for binarity in this object.  Objects that looked binary or had a close star in the field on pair-subtracted images were also observed with the MKO $J$ and/or $K_s$ filters.  Details of the observing log can be found in Table~\ref{tab_obs}.

The images were reduced in a standard fashion using IDL scripts.  Because of the small FOV of the images and the scarcity of objects on the frames, we used a simple ``A$-$B'' method for the sky subtraction, where ``A'' is the target image to be reduced and ``B'' is the dithered frame of the target either immediately after or before ``A,'' depending on where ``A'' was in the full dither sequence.  This method produced a much better sky subtraction compared to the median of all dithered observations of a target.  The presence of negative stars from the offset sky frame had no impact on our ability to detect companions because they were significantly offset (4-5$\arcsec$) from the target in most cases.  One of the targets (WISE 1653+4444) had a tip-tilt reference star near the edge of the field steering mirror field-of-view and required small 1$\arcsec$ nods to keep the lock on the tip-tilt star.  Even these small offsets are well beyond the PSF of the target and only had a minor impact on the sensitivity of companion detection at that radius (see below).  The sky-subtracted frames were then divided by a dome flat.  The images were shifted to align the target to a common location and the stack was median averaged to create the final mosaics (see Figures~\ref{fig_mosaic}, \ref{fig_w0750mosaic}, \ref{fig_1841mosaic}, and \ref{fig_0458mosaic}) .

\section{Results\label{results}}
\subsection{Sources With No Apparent Companions}
The shape of the target point spread function (PSF) is heavily dependent on the observing conditions (seeing and airmass), returned power of the LGS, and the brightness and location (relative to the target) of the tip-tilt reference star.  In general, reference stars that are within about 20$\arcsec$ of the target produce fairly circular PSFs, whereas reference stars between $\sim$20-60$\arcsec$ can produce elliptical PSFs aligned in the direction of the reference star (see WISE 1653+4444 in Figure~\ref{fig_mosaic}).  Thus, it is often difficult to identify very close binaries based solely on the shape of the PSF when the target is the only source in the very small field.  It is possible to select ``PSF'' targets with reference star properties similar to that of the science target in order to approximate the PSF of a single star in our field of interest.  Unfortunately, we were unable to obtain observations of PSF stars for the observations presented here.  Nonetheless, given the smooth slopes of our targets' PSFs, we are confident in our ability to identify bright companions beyond $\approx$0$\farcs$3.

Seven of our targets have PSFs that are consistent with unresolved point sources.  We utilize two quantitative methods to compute the magnitude limits for companions close to the target and describe both in the following sections.

\subsubsection{PSF Planting}
Our first method to compute sensitivity involves planting fake companions on the image at various radii from the target, and determining if those companions can be detected using automated software.  We first make a 257$\times$257 pixel subimage of the field with the location of the target's centroid in pixel position (129,129).  Assuming that the PSF of the target is rotationally symmetric, we rotate the image 180$\degr$ and subtract it from an unrotated copy of the image.  We shift the rotated image systematically around a 7$\times$7 pixel grid ($\pm$3 pixels in X and Y about the center) to find the optimal offset whose subtraction produces the lowest residuals.

The synthetic companion PSF is simply a scaled version of the target's PSF that is added to the residual image at 20 different radii from the central pixel (distances range from 5 to 100 pixels in steps of 5 pixels).  At each radius, we step through 20 companion magnitudes (from 0.25 to 5.0 magnitudes {\it fainter} than the target) and plant them at 100 random position angles (i.e. there are 40,000 companions with different radius+magnitude+position angle combinations).  We use SExtractor \citep{bertin1996} to search for star-like sources in the field and are specifically interested in the code's ability to find the planted companion\footnote{We largely use the default parameters (eg., {\tt DETECT\_MINAREA}=5, {\tt DETECT\_THRESH}=1.5, and {\tt ANALYSIS\_THRESH}=1.5) for object extractions.}.  We define a ``found'' source as one whose SExtractor {\tt X\_BEST} and {\tt Y\_BEST} coordinates are within 3 pixels of the location of the planted object.  The limiting magnitude for detection is the faintest magnitude at which SExtractor finds more than 50\% of the planted PSFs at a given radial distance from the central pixel.  The 50\% limit is somewhat arbitrarily chosen, but is based on SExtractor's inability to find the planted companion even in cases in which the companion is clearly visible.  Requiring that 90-100\% of the planted companions to be found results in magnitudes limits that are much too bright compared to limits finding the companions visually.  A threshold of 50\% provides limits that are more reasonable, though are perhaps still too bright.

Figure~\ref{fig_psf_limits} shows the sensitivity obtained using PSF planting.  The results of this technique internal to a radius of 0$\farcs$2 are quite unreliable.   Residuals from the rotation-subtracted target can distort the companion's PSF in such a way that SExtractor has difficulty finding the companion.  Beyond 0$\farcs$2 the residuals are much more uniform and the companion's PSFs do not change significantly at the various position angles.  This is readily apparent in the relative flatness in the sensitivity beyond 0$\farcs$2.  

\subsubsection{Standard Deviation of the Background}
Our second method for computing the sensitivities appears to be more reliable than the PSF-planting technique.  For this method we create a set of $\approx$100 annuli with 1-pixel widths centered on the target, in the mosaic image, and running out to 100 pixels ($\approx1\arcsec$ for the narrow camera and 4$\arcsec$ for the wide camera).  We calculate the standard deviation of the pixel values in each annulus ($\sigma_{an}$) and compute the flux an object would have if it is found in the annulus and has a measurement error of 0.33 mag (3-$\sigma$ measurement).   The flux of the companion, $F$, is calculated by starting with the standard equations for the magnitude error ($\sigma_m$):

\begin{equation}
\sigma_m = 1.0857 * \sigma_F / F
\end{equation}
\begin{equation}
\sigma_F = \sqrt[]{(A*\sigma_{an} ^2)^2 + (A^2*\sigma_{an}^2/N_{an})^2 + (F/G)^2}
\end{equation}

\noindent where $\sigma_m$ is the magnitude error for detection limits (0.33 mag), $G$ is the NIRC2 gain (4 e$^-$/DN), $A$ is the area of the aperture used for the hypothetical measurement of the companion ($\pi$*[10pix]$^2$ for the narrow camera observations and $\pi$*[5pix]$^2$ for the wide camera observations),  and $N_{an}$ is the number of pixels in the annulus.  Combining Equations 1 and 2 and solving for $F$ gives

\begin{equation}
F = \frac{1.0857^2/(\sigma_m^2 * G) + \sqrt[]{(1.0857^2/(\sigma_m^2*G))^2 +  4*(A*\sigma_{an}^2 + A^2*\sigma_{an}^2/N_{an})*1.0857^2/\sigma_m^2}}{2},
\end{equation}

\noindent This flux is converted to a magnitude and the magnitude of the target as measured on the frame is subtracted from it to obtain the magnitude difference (Figure~\ref{fig_limits}).    In general, we are able to reach limiting magnitudes of $\approx$22 mag in our exposures for companions with separations greater than 0$\farcs$3.  For an assumed distance of 10pc (projected separation of 3 AU) and age of 1 Gyr, this corresponds to a mass and effective temperature of 6 $M_{\rm Jup}$ and 350 K, respectively \citep{bsl2003}.

\subsection{Sources With Binary PSFs}
Two of our targets have PSFs with structure suggestive of them being multiple objects.  In the following sections we discuss each of these targets in more detail.

\subsubsection{WISE 1841+7000}
With a spectral type of T5, WISE 1841+7000 is the warmest object in our sample.  Both the individual images and the final mosaics (Figure~\ref{fig_1841mosaic}) reveal a dual-peaked PSF with a separation of $\approx$8 pixels (0$\farcs$08).    The close proximity of the sources complicates the relative photometry because the PSFs overlap considerably.  We used the rotate and subtract method described in \citet{looper2008} to reduce the contribution of the adjacent object when measuring the flux of a source.

Analysis of the images are further complicated by the quality of the PSFs.  The reference star is located $\approx$46$\arcsec$ from the target and the observations were taken at an airmass of 1.61.  Consequently, the PSFs have a fuzzy, elliptical shape, which is clearly evident on the mosaicked images and even more so on the companion-subtracted images (Figure~\ref{fig_1841_rotsub}).  Using a circular aperture for the photometry would have been inefficient as it would have either included large amounts of sky or chopped off flux from the source.  Therefore, an elliptical aperture was used to compute the flux of the source (circular annuli were still used for the sky computation).  The magnitudes of the sources are highly dependent upon the quality of the companion's subtraction which, in turn, depends primarily on the pixel location about which the image is rotated.  For a given object, this rotation axis was chosen by computing a Gaussian centroid for the object.  Because this centroid can be affected by the companion and to assign errors to our relative magnitudes, we compute the instrumental magnitudes for rotation axes distributed over a 3$\times$3-pixel grid centered on the best rotation axis.  The final magnitudes are the average of the 9 magnitudes, and the errors are the standard deviations of the 9 magnitudes.  The resultant relative photometry is shown in Table~\ref{tab_binprop}, where we refer to the brighter $J$ component to the West as WISE 1841+7000A and the fainter component $J$ component to the East as WISE 1841+7000B (see below). 

The differential $H$ and $K_s$ magnitudes are consistent with equal brightness components ($\Delta H$=0.02$\pm$0.12; $\Delta K_s$=0.10$\pm$0.09), and the magnitude difference in $J$ ($\Delta J$=0.33$\pm$0.17 mag) is only significant at the 2-$\sigma$ level.  There are only two other known binaries that are nearly equal mass with spectral types similar to WISE 1841+7000AB: 2MASS 1534$-$2952AB \citep[T5+T5.5;][]{2003ApJ...586..512B,2008ApJ...689..436L} and SDSS 0926+5847AB \citep[T4+T4;][]{2006ApJS..166..585B}.  Although 2MASS 1534$-$2952AB has NIRC2 observations similar to our own, the near-IR colors for that system \citep{2008ApJ...689..436L} are not similar to those of WISE 1841+7000AB.  The angular separation of the 2MASS 1534$-$2952AB system is $\approx$3 times larger than the separation for WISE 1841+7000AB and did not require any special PSF-subtraction techniques to isolate the components.  Consequently, the photometry for that system is presumably more robust than the photometry presented here.  

SDSS 0926+5846AB is similar to WISE 1841+7000AB in that those components are also nearly equal brightness. The $HST$ F110W and W170M differences for SDSS 0926+5846AB are $\lesssim$2-$\sigma$ results (0.4$\pm$0.2 and 0.4$\pm$0.3, respectively).  As with 2MASS 1534$-$2952AB, the colors of the SDSS 0926+5846AB components are quite unlike those of WISE 1841+7000AB.  Neither one of these other binaries is like WISE 1841+7000AB, but there are two non-astrophysical reasons why this could be: our subtraction technique for removing the flux from the companion may not be ideal since it is possible to either over- or under-subtract the companion and the MKO filters are quite different from the ones used on $HST$.  Higher quality follow-up observations, particularly with HST which has a more stable PSF than NIRC with LGS-AO, will be able to determine if this binary is indeed anomalous compared to similar binaries.
  
We use the spectral type--absolute magnitude relationship parameterized in \citet{marocco2010} to estimate the distance to W1841+7000AB.  By fitting the absolute magnitudes of L and T dwarfs from L0 to T9, \citet{marocco2010} are able to derive the relationship between absolute magnitude and spectral type for the $J$, $H$, and $K_s$ MKO filters.   Since we do not have a calibrated $K_s$ magnitude for the WISE 1841+7000AB composite, we can only compute absolute magnitudes for $J$ and $H$.  We assume that the spectral type for both components of the system are equal (T5), since the flux ratios at these bandpasses are nearly 1.  We use both fits (including and excluding suspected binaries) to estimate a $J$-band distance of 38.3$\pm$4.8 pc and an $H$-band distance of 42.1$\pm$4.9 pc, for a average distance of 40.2$\pm$4.9 pc and a projected component separation of 2.8$\pm$0.7AU.  For an assumed age of 1 Gyr, those objects would have a mass of $\approx$60 times that of Jupiter \citep{chabrier2000}, corresponding to an orbital period of $\sim$11 years.  

With only a single epoch showing the resolved components, the validity of calling this a binary system comes into question.  Fortunately, the ancillary data that exists for this object rule out the fainter source as a background object.  WISE 1841+7000 has a proper motion of $\approx$0.5 $\arcsec$/yr (Kirkpatrick et al., submitted) and is visible in 2MASS.  The 2MASS observation was obtained 12 years prior to the NIRC2 images, and the brown dwarf has moved more than 6$\arcsec$ between these epochs (the proper motion is confirmed by multiple epochs from WISE and $Spitzer$, so we are confident that the 2MASS source is WISE 1841+7000).  If one of these sources was a background object, then it should also be visible in the 2MASS image at approximately the same brightness as the brown dwarf.  The lack of any object in the 2MASS image at the WISE-derived position of the brown dwarf indicates that the sources seen in the NIRC2 image are co-moving and gravitationally bound.

\subsubsection{WISE 0458+6434}
WISE 0458+6434 is the first ultracool brown dwarf found using WISE data \citep{mainzer2011} and has a spectral type of T8.5 (Cushing et al., in prep.).    This object was observed at the start of the night, just before it reached its maximum allowed airmass for LGS AO observations (2 airmasses).   The performance of the LGS AO system is adversely impacted at such a large airmass, resulting in poor AO correction and very fuzzy images (Figure~\ref{fig_0458mosaic}).  However, our first observations at $H$ clearly showed the object was a binary.  Since we had limited time to observe it before it hit the maximum airmass, we obtained follow-up images in $J$-band only.  

Given the poor quality of the images, the computation of the component separation and position angle proved to be more difficult than usual for LGS AO data.  In fact, the images are sufficiently fuzzy that a centroid location for the companion could not be found for either mosaic.  In order to compute the separation, we made several measurements of the apparent central peaks of the components as judged by-eye.  We then found the average and standard deviation of the different measurements and present them in Table~{\ref{tab_binprop}.  

The photometry is a little more straightforward than for WISE 1841+7000AB because the angular separation is large enough ($\approx$50 pixels) that special star subtraction techniques are not needed to obtain good photometry.  The main difficulty comes because there is no centroid location for either component.  To obtain the photometry, we start with what appears (by eye) to be the central location of each component.  We compute the photometry of the objects with a circular aperture centered on that location.  We then allow the center of the aperture to vary in both X and Y by $\pm$3 pixels (integer pixel steps) in order to obtain a good sample of other possible object ``centroids.''  The instrumental magnitudes are the average of the 49 measurements and the errors are the average of the formal measurement errors added in quadrature with the standard deviation of the instrumental magnitudes.  The differential magnitudes are listed in Table~{\ref{tab_binprop}, as are the magnitudes of the individual components.  

It's important to note that the composite near-IR magnitudes for this source in Table~\ref{tab_targs} are calibrated with 2MASS stars and using 2MASS-like filters.  The filters in NIRC2 are on the MKO system so the magnitude differences measured on these images might not represent the magnitude differences as seen with 2MASS filters, particularly for $J$, which is considerably wider in the 2MASS system.  To correct for the difference in the filter systems, we derived the $J$ and $H$ offsets between MKO and 2MASS by convolving near-IR spectra we have obtained for several late T dwarfs (spectral types $\ge$T8) with the filter profiles for the 2MASS and MKO $J$ and $H$ filters.  The objects and offsets are provided in Table~\ref{tab_offsets}.  The average magnitude differences for this ensemble ($J_{\rm 2MASS} - J_{\rm MKO}$ = $+$0.339$\pm$0.008 mag and $H_{\rm 2MASS} - H_{\rm MKO}$ = $-$0.039$\pm$0.007 mag) are consistent with the offsets between 2MASS and MKO derived by  \citet{stephens2004} for late T dwarfs.  We apply these offsets to the composite magnitudes in Table~\ref{tab_targs} before using our measured magnitude differences to compute the individual magnitudes.

We use the same method described above for WISE 1841+7000AB to estimate the distance to WISE 0458+6434AB.  We assume that the flux and features in the composite spectrum are dominated by the brighter component (WISE 0458+6434A), and therefore assign the composite spectral type of T8.5 to that source.  We find that the distance computed based on the $J$ and $H$ magnitudes of WISE 0458+6434A are consistent with each other (10.6$\pm$1.3 pc and 10.4$\pm$1.4 pc for $J$ and $H$, respectively) and take the average of these values as the distance to the system (10.5$\pm$1.4 pc).  

To estimate the spectral type of WISE 0458+6434B, we use the distance derived above to compute the absolute $J$ magnitude for WISE 0458+6434B.  At M$_J$= 18.4$\pm$0.3, it is consistent with the absolute $J$ magnitude of UGPS 0722$-$0540 \citep[18.5$\pm$0.2;][]{lucas2010} and suggests that both objects have similar effective temperatures and spectral types.  This would imply that both UGPS 0722$-$0540 and WISE 0458+6434B have spectral types of T9 (Cushing et al., in prep.)

We adopt a spectrophotometric distance of 10.5$\pm$1.4 pc for WISE 0458+6434A and B and use the absolute magnitudes to compare them to late-type T dwarfs with measured parallaxes (Figure~\ref{fig_0458tcolors}).  WISE 0458+6434AB is consistent with other late-type T dwarfs and inhabits the regime occupied by T8.5 and T9 dwarfs, confirming that both the component colors computed here and the estimated distance are reasonable.  Comparison to the models of  \citet{bsl2003} and \citet{hb2007} shows that WISE 0458+6434A has an effective temperature $\sim$600 K ($M\approx M_{\rm Jup}$ at 1Gyr) and WISE 0458+6434B has an effective temperature $\sim$500 K (consistent with UGPS 0722$-$0540 and equivalent to a 10 $M_{\rm Jup}$ object at 1Gyr).  A measured parallax and resolved spectroscopy will permit a more robust comparison to the models and other T dwarfs, but it is clear from just this crude analysis that these objects are among the coolest known brown dwarfs.  With a projected physical separation of $\approx$5 AU and an orbital period $\sim$70 years (assuming a total system mass of 25 $M_{\rm Jup}$), this system will be an important benchmark for late-type T dwarfs as the orbital motion is followed over the coming decades.

The proper motion of WISE 0458+6434 is only 0.2 $\arcsec$/yr and the brown dwarf is too faint to have been detected by 2MASS (Kirkpatrick et al., submitted).  Furthermore, the 2 month separation between this NIRC2 data and the WISE discovery observations is not nearly long enough for the brown dwarf to have moved significantly far from a background object and allow both objects to be visible in the WISE image.  Even the year-long baseline between the WISE observations and those from IRAC $Spitzer$ (Kirkpatrick et al., submitted), provides for motions less than the resolutions of those respective instruments.  Another epoch of high resolution imaging should be able to confirm these objects as a proper motion pair.  

As mentioned above, the WISE W1-W2 colors of cool brown dwarfs are very red ($>$2).  WISE 0458+6434 has a W1-W2 color of 3.4 (Table~\ref{tab_targs}), which is consistent with similarly-typed brown dwarfs (Kirkpatrick et al., submitted).  In addition, the near-IR spectra \citep{mainzer2011} do not show any signs of contamination by a non-brown dwarf object.  The fainter object contributes $\sim$30\% of the total light in the system and its contribution, if it did not have a similar spectral type to the brighter component, would be easily discernible.  The lack of any peculiarity in both the composite photometry and spectroscopy, coupled with the strong resemblance of the fainter source to the confirmed T9 UGPS 0722$-$0540, leads us to conclude that both objects are brown dwarfs.

\section{Conclusions}
We have observed nine new WISE brown dwarfs with the Keck II LGS-AO system in an attempt to discover ultracool companions.  Seven of our sources do not show evidence of binarity with limits of $\Delta H \approx3-5$ mag at 1$\arcsec$.  WISE 1841+7000AB is a T5 binary with nearly equal brightness components in $H$ and $K_s$, but a $\Delta J$ of 0.33$\pm$0.17 mag.  Its photometry is unlike other binaries with similarly typed components; follow-up work will be important for determining if these differences point to real astrophysical differences.  The  other binary, WISE 0458+6434AB, is noteworthy because, with a composite spectral type of T8.5, it is the one of the latest-type brown dwarf binaries known.  The companion is $\approx$1 mag fainter than the primary and has similar properties to UGPS 0722$-$0540 \citep{lucas2010}.  This binary represents an important benchmark system for brown dwarf atmosphere models and will benefit greatly from resolved spectroscopy and astrometric monitoring.

\acknowledgements
The authors acknowledge telescope operators Julie Rivera, Heather Hershley, and Gary Punawai,  
and instrument specialists Al Conrad, Marc Kassis, and Scott Dahm at Keck,
for their assistance during the
observations.  
This publication makes use of data products from the Wide-field Infrared Survey Explorer, which is a joint project of the University of California, Los Angeles, and the Jet Propulsion Laboratory/California Institute of Technology, funded by the National Aeronautics and Space Administration.  This research was supported in part by an appointment to the NASA Postdoctoral Program at the Jet Propulsion Laboratory, administered by Oak Ridge Associated Universities through a contract with NASA.  Data presented herein were obtained at the W. M. Keck Observatory from telescope time allocated
to the National Aeronautics and Space Administration through the agency's scientific partnership
with the California Institute of Technology and the University of California. 
The authors recognize and acknowledge the 
very significant cultural role and reverence that 
the summit of Mauna Kea has always had within the 
indigenous Hawaiian community.  We are most fortunate 
to have the opportunity to conduct observations from this mountain.

Facilities: \facility{Keck~(NIRC2,LGS)}

\clearpage

\begin{figure}
\epsscale{0.25}\plotone{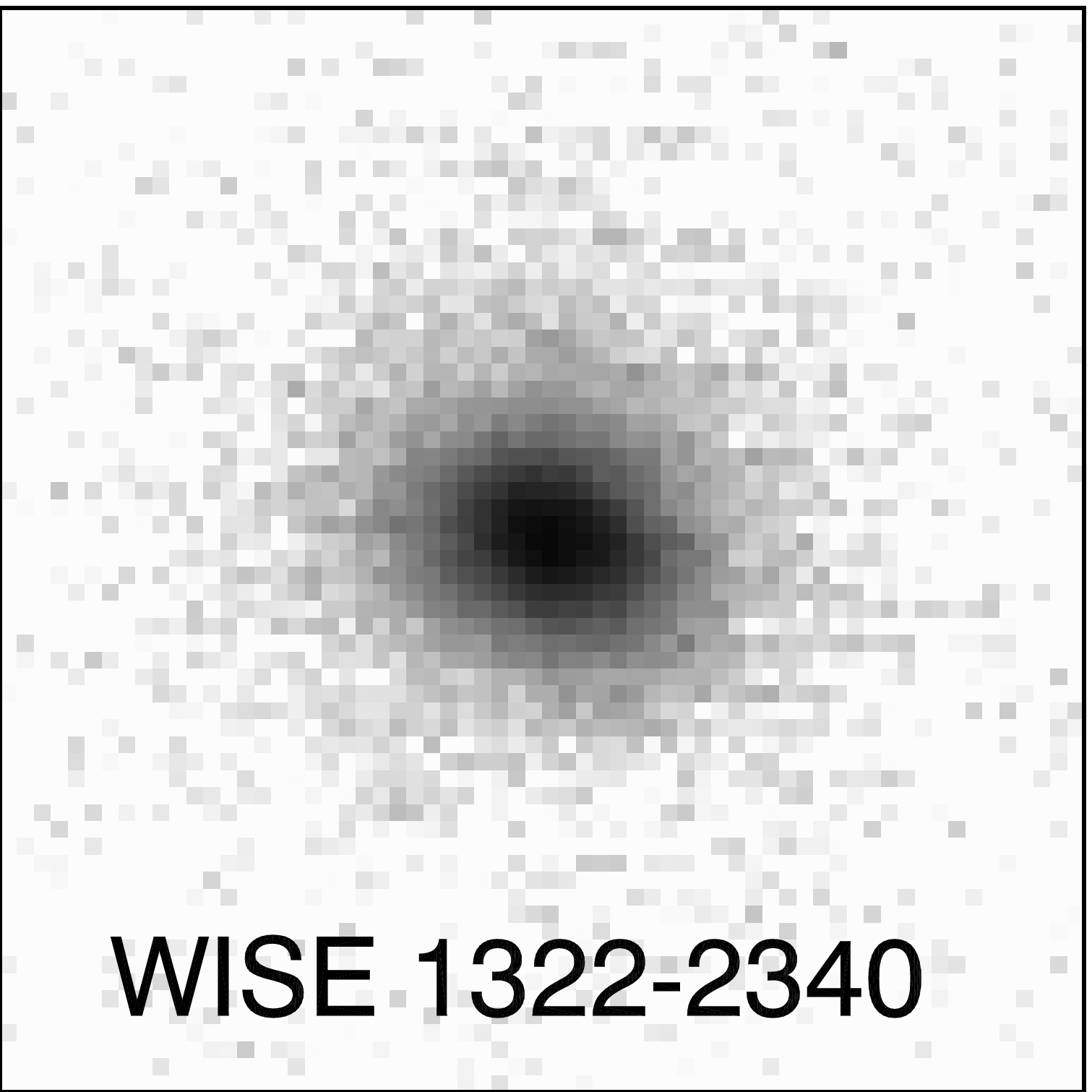}
\epsscale{0.25}\plotone{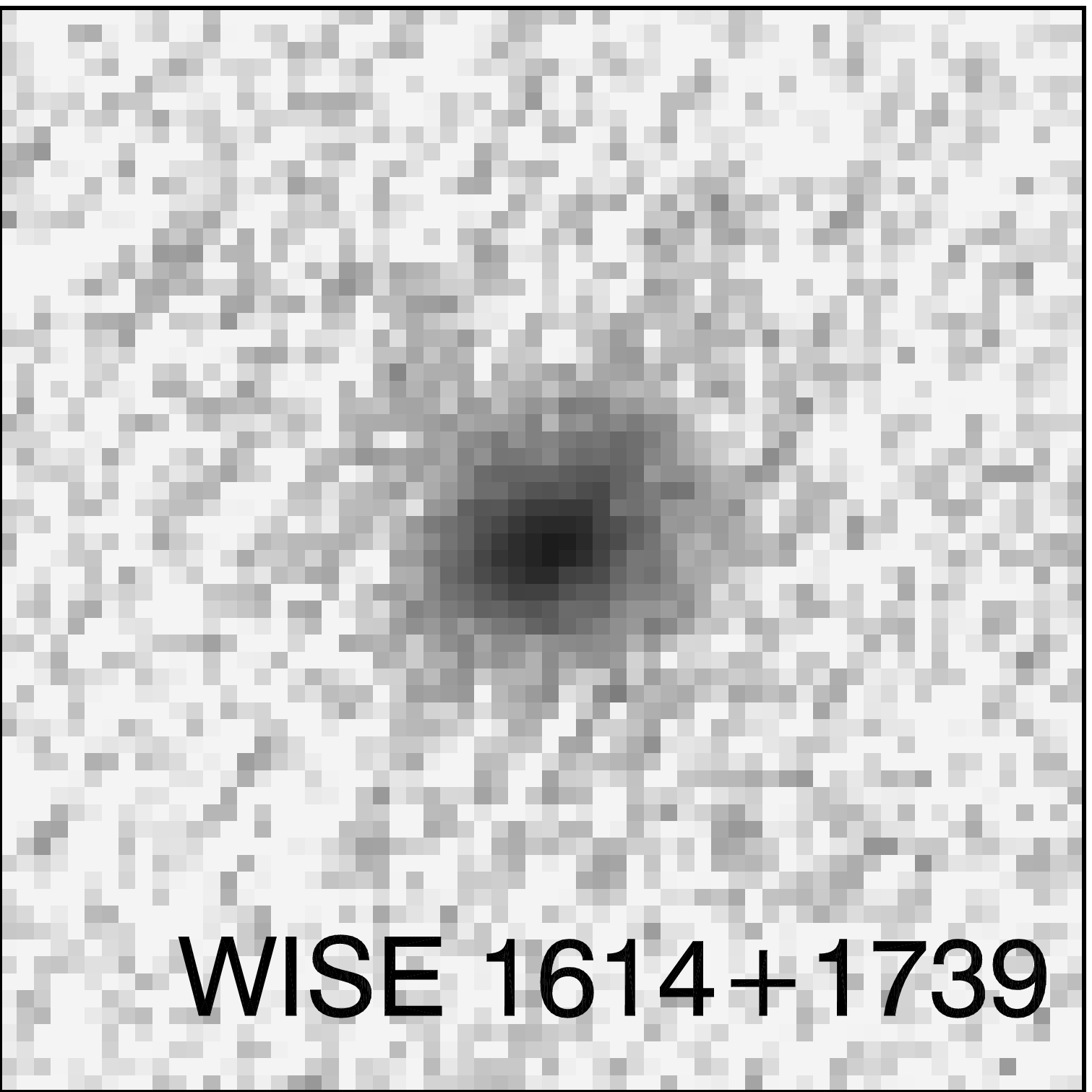}
\epsscale{0.25}\plotone{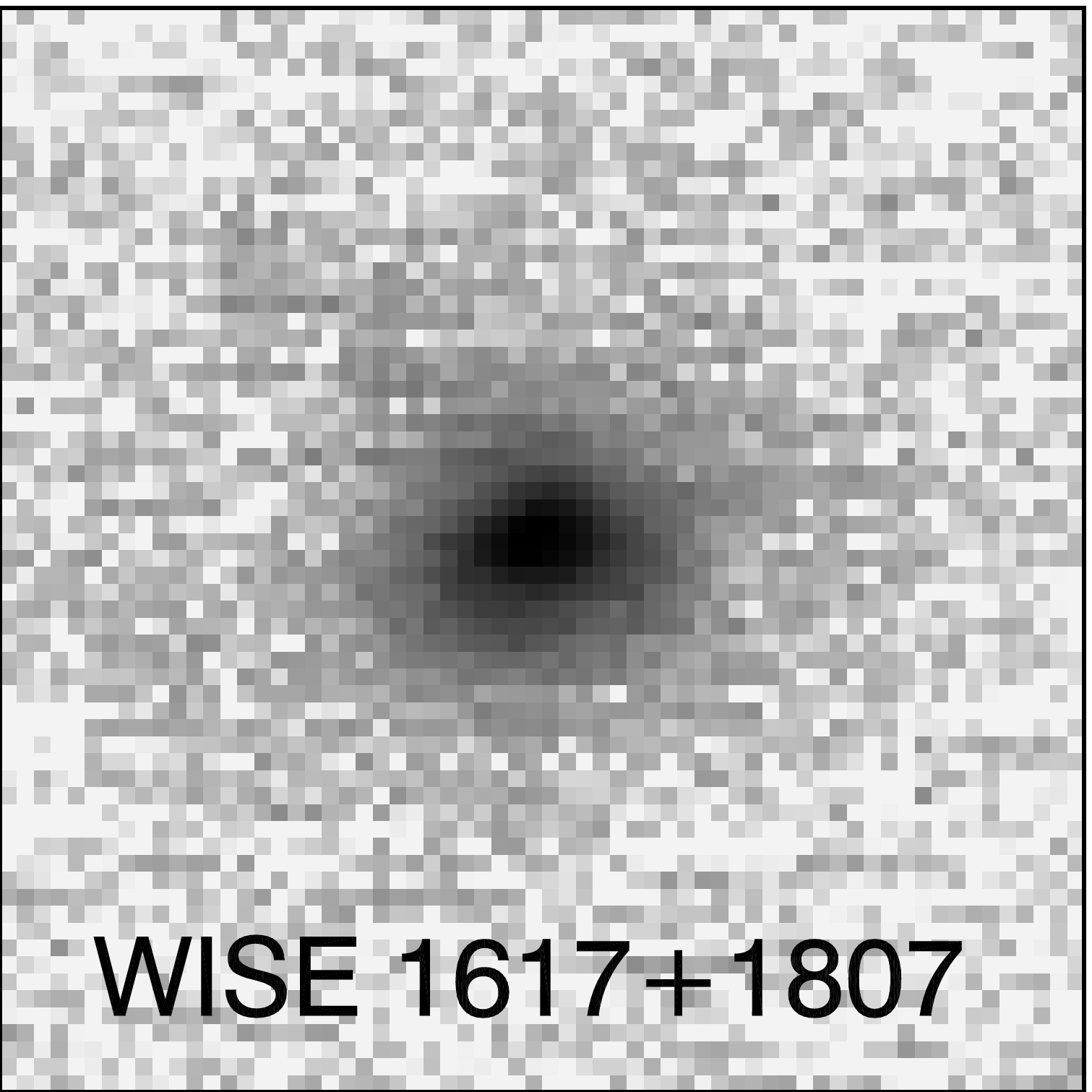}
\epsscale{0.25}\plotone{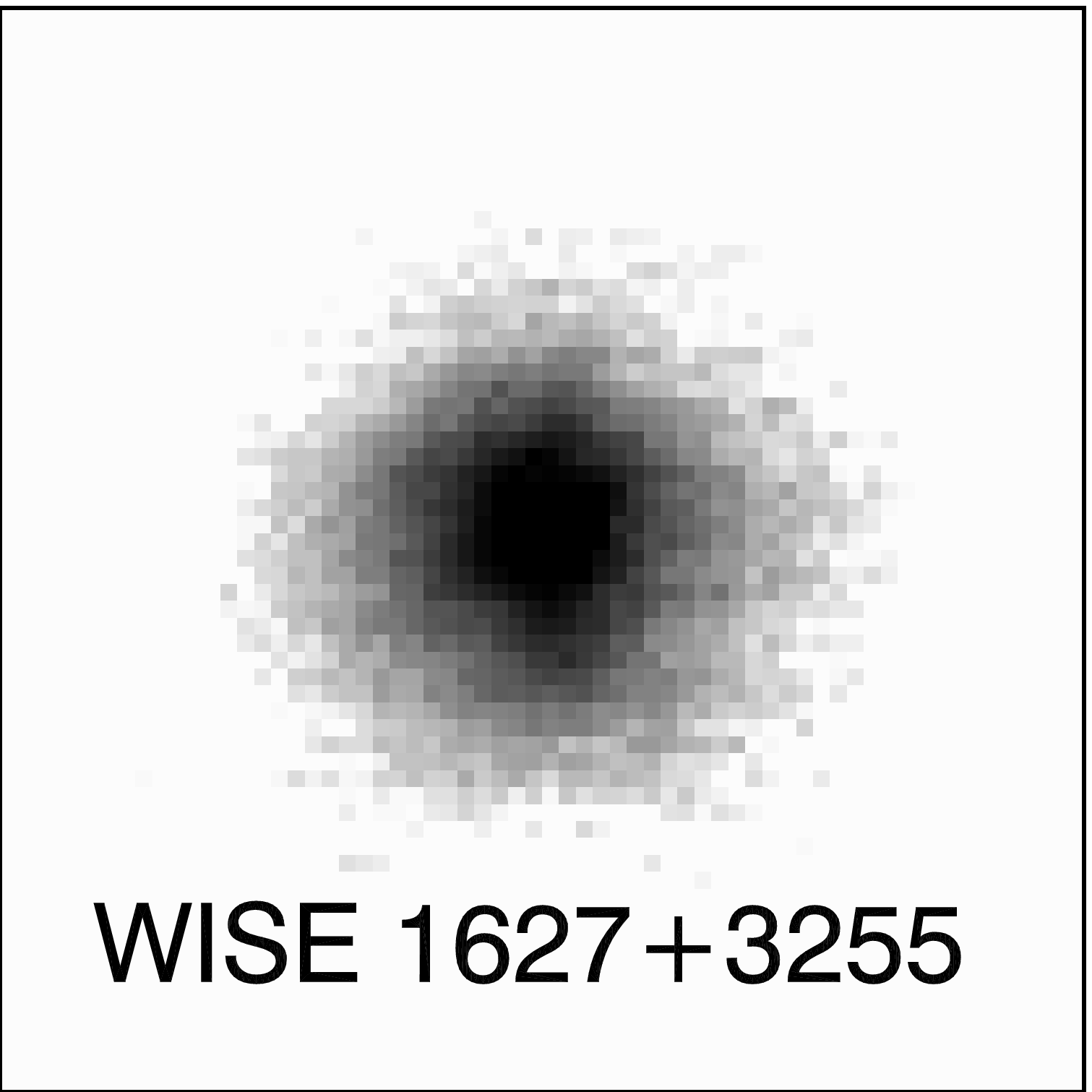}
\epsscale{0.25}\plotone{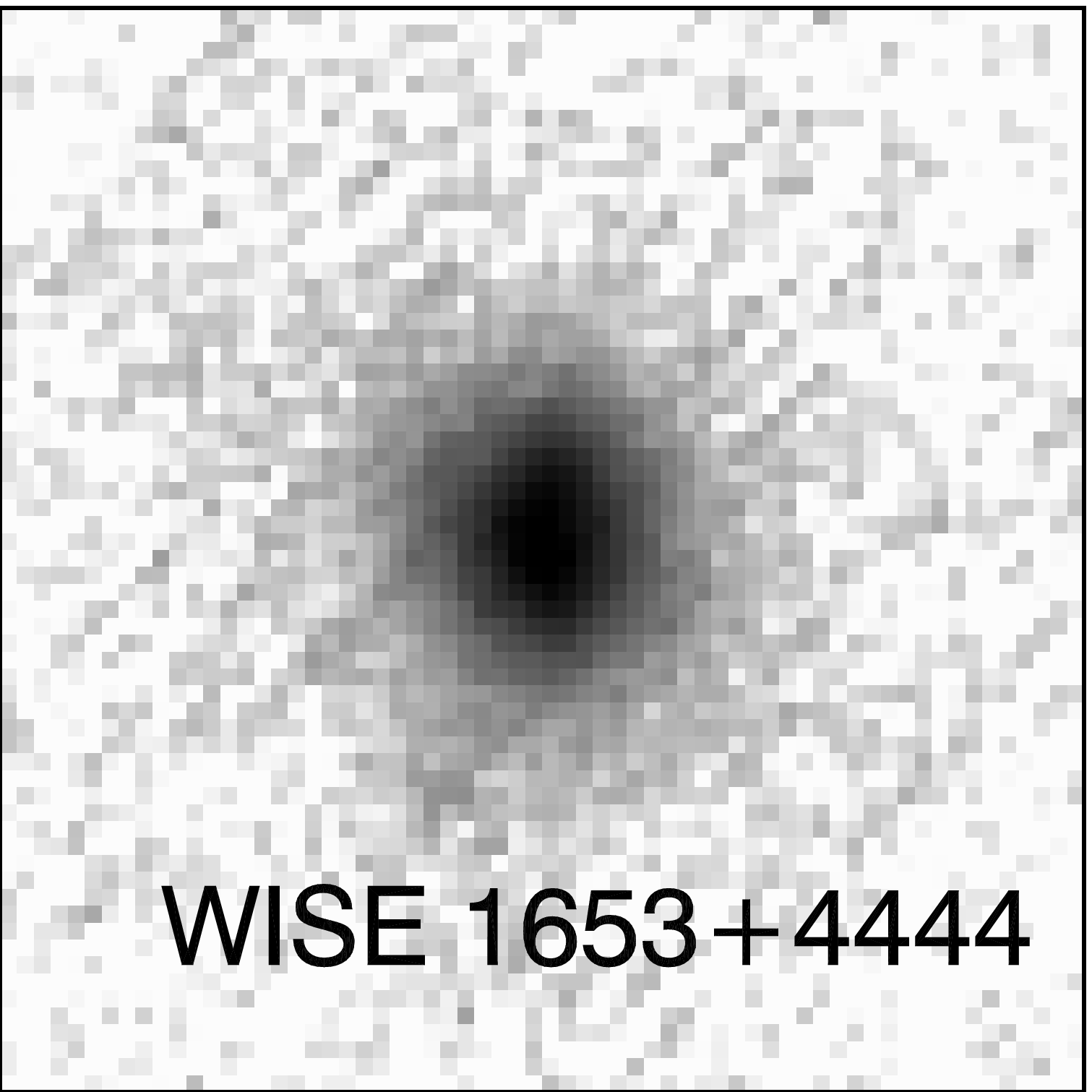}
\epsscale{0.25}\plotone{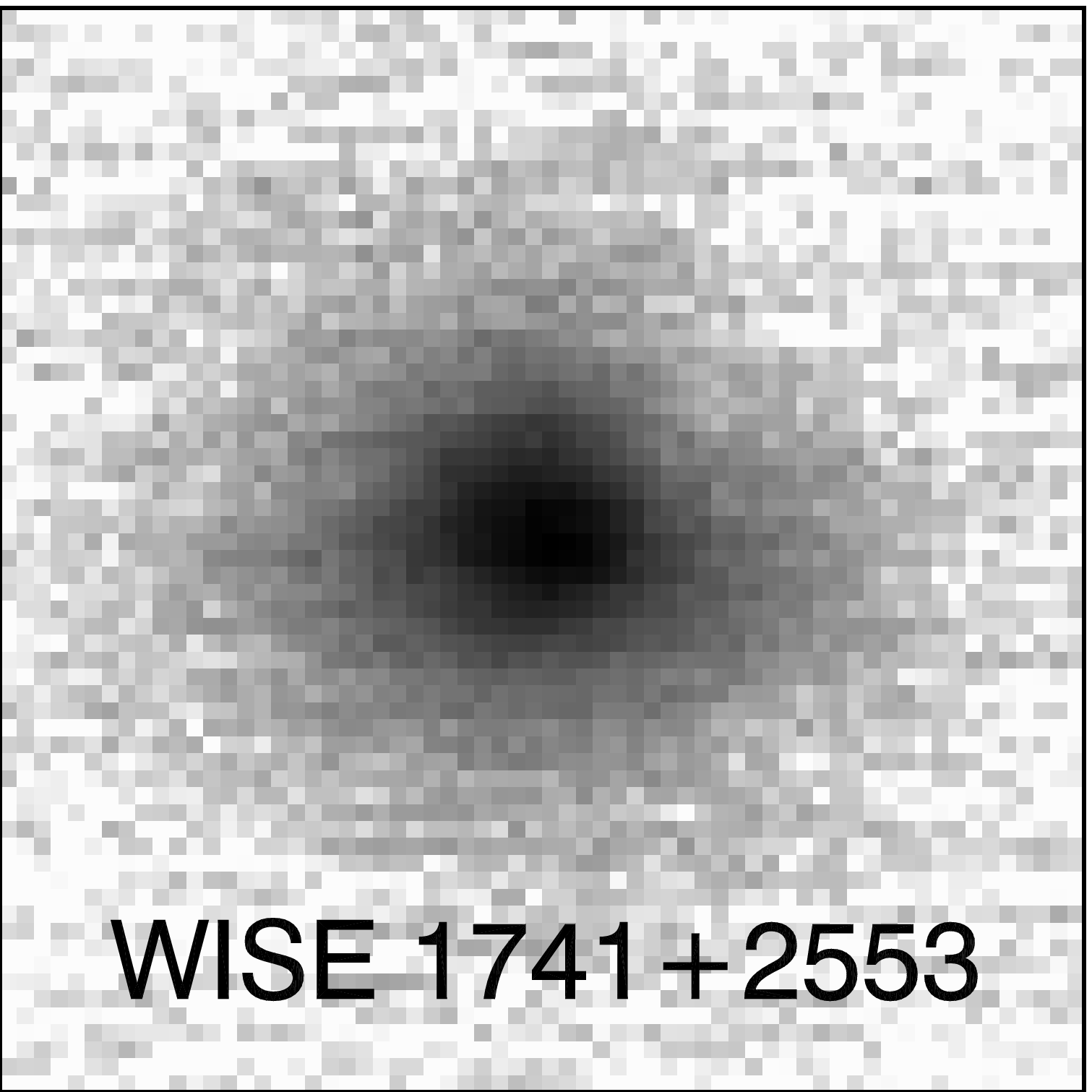}
\caption{$H$-band LGS-AO images for six of our apparently single targets.  The images are $\approx$0$\farcs$6 on a side with North up and East to the left. \label{fig_mosaic}}
\end{figure}

\begin{figure}
\epsscale{1.0}\plottwo{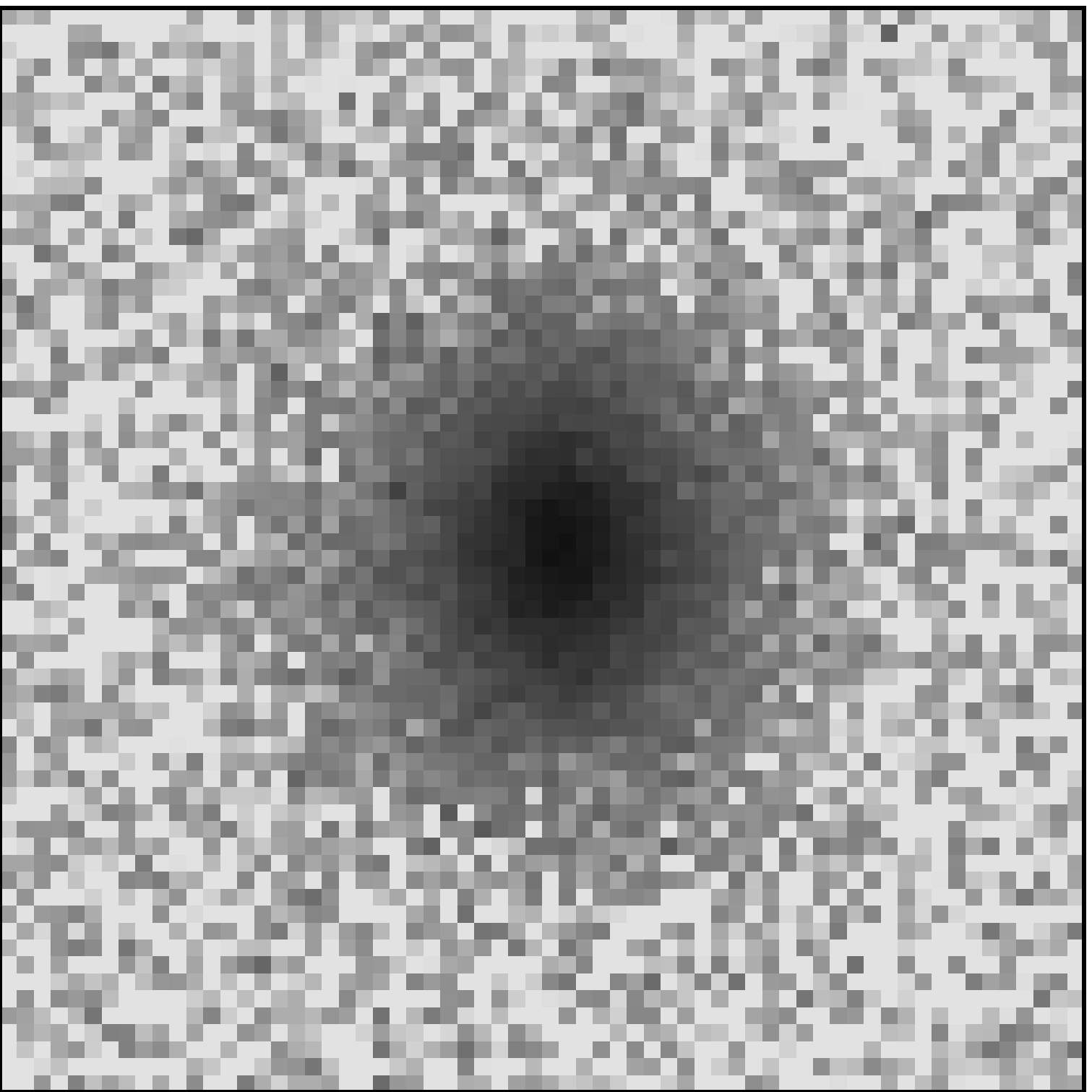}{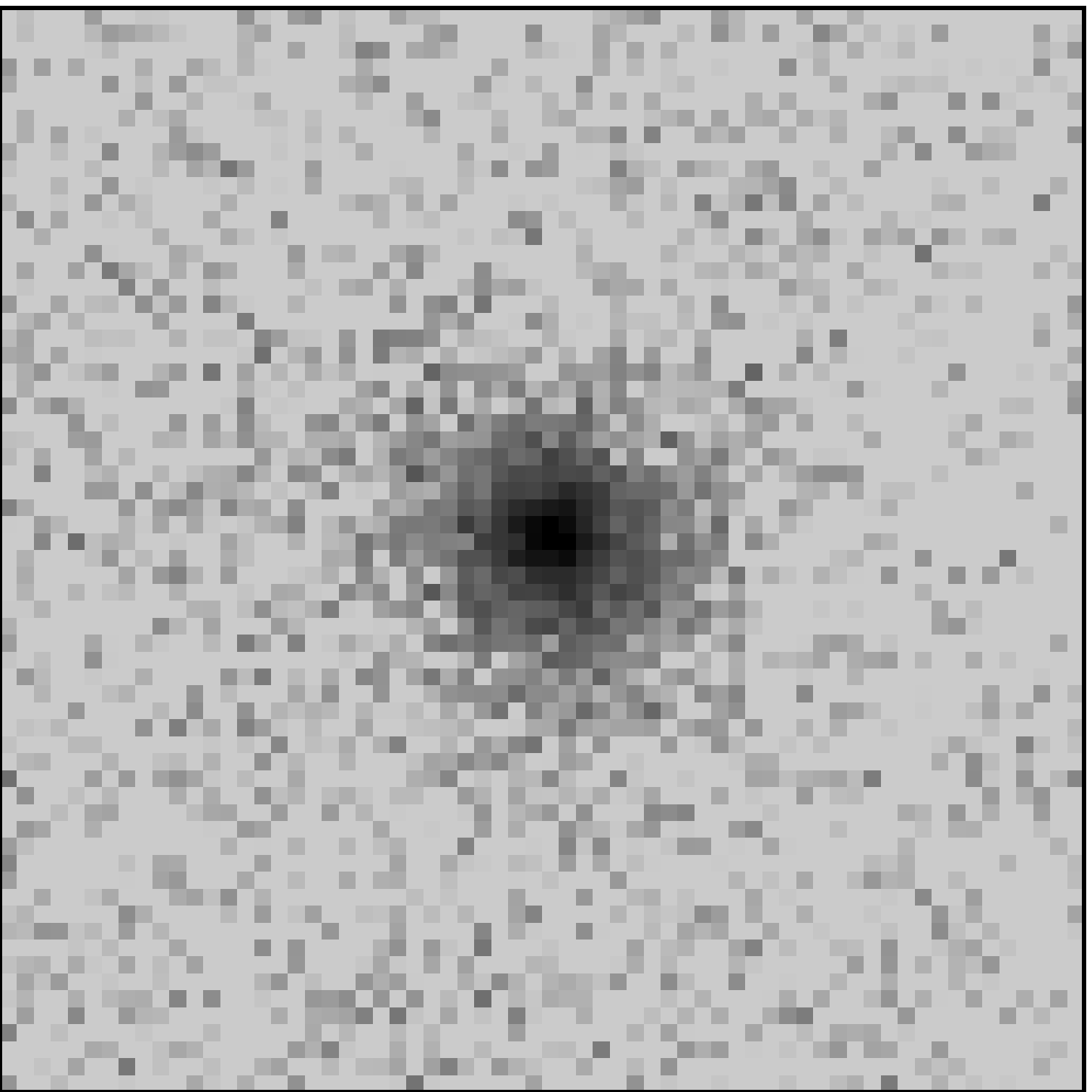}
\caption{$J$- (left) and $H$-band (right) LGS-AO images of WISE 0750+2527.  The images are $\approx$2.6$\arcsec$ on a side with North up and East to the left and were obtained with the NIRC2 wide camera setting.  The PSF shows no obvious signs of binarity.\label{fig_w0750mosaic}}
\end{figure}

\begin{figure}
\epsscale{1.0}
\plotone{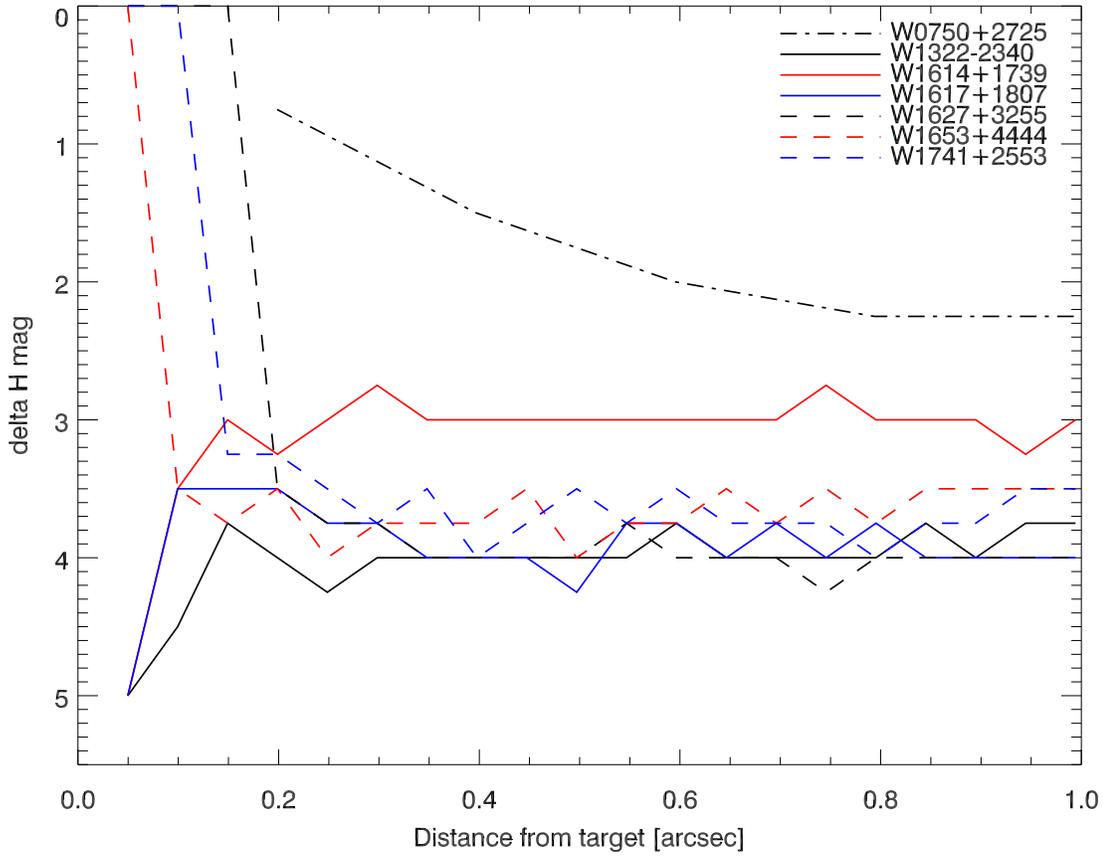}
\caption{Magnitude difference limits for the detection of a faint companion using the PSF-planting technique.  Results within 0$\farcs$2 are not reliable because the planted PSF is contaminated by the residuals from the rotation-subtracted target. The apparent magnitude detection limits vary with each source, but are in the range $H \approx$19--22 mag.\label{fig_psf_limits}}
\end{figure}

\begin{figure}
\epsscale{1.0}
\plotone{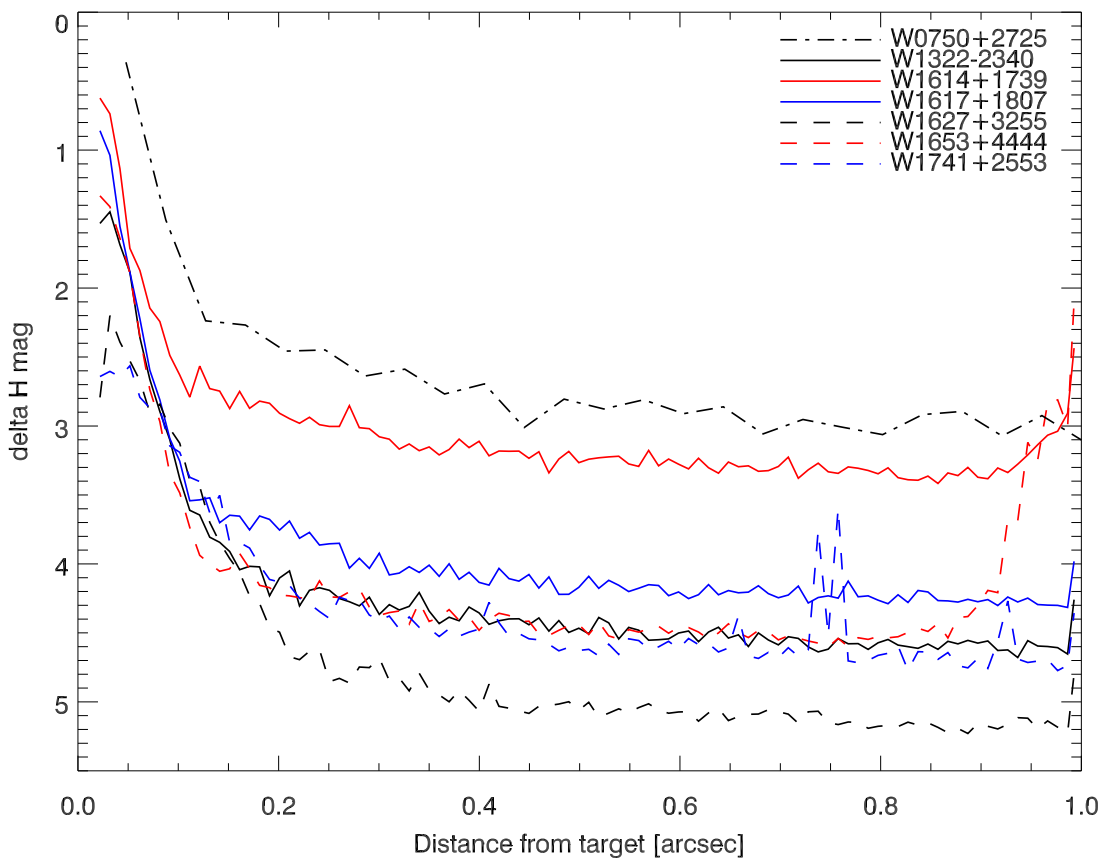}
\caption{Magnitude difference limits for the detection of faint companion using the standard deviation of the background technique. The apparent magnitude detection limits vary with each source, but are in the range $H \approx$20--22 mag.\label{fig_limits}}
\end{figure} 

\begin{figure}
\epsscale{0.25}
\plotone{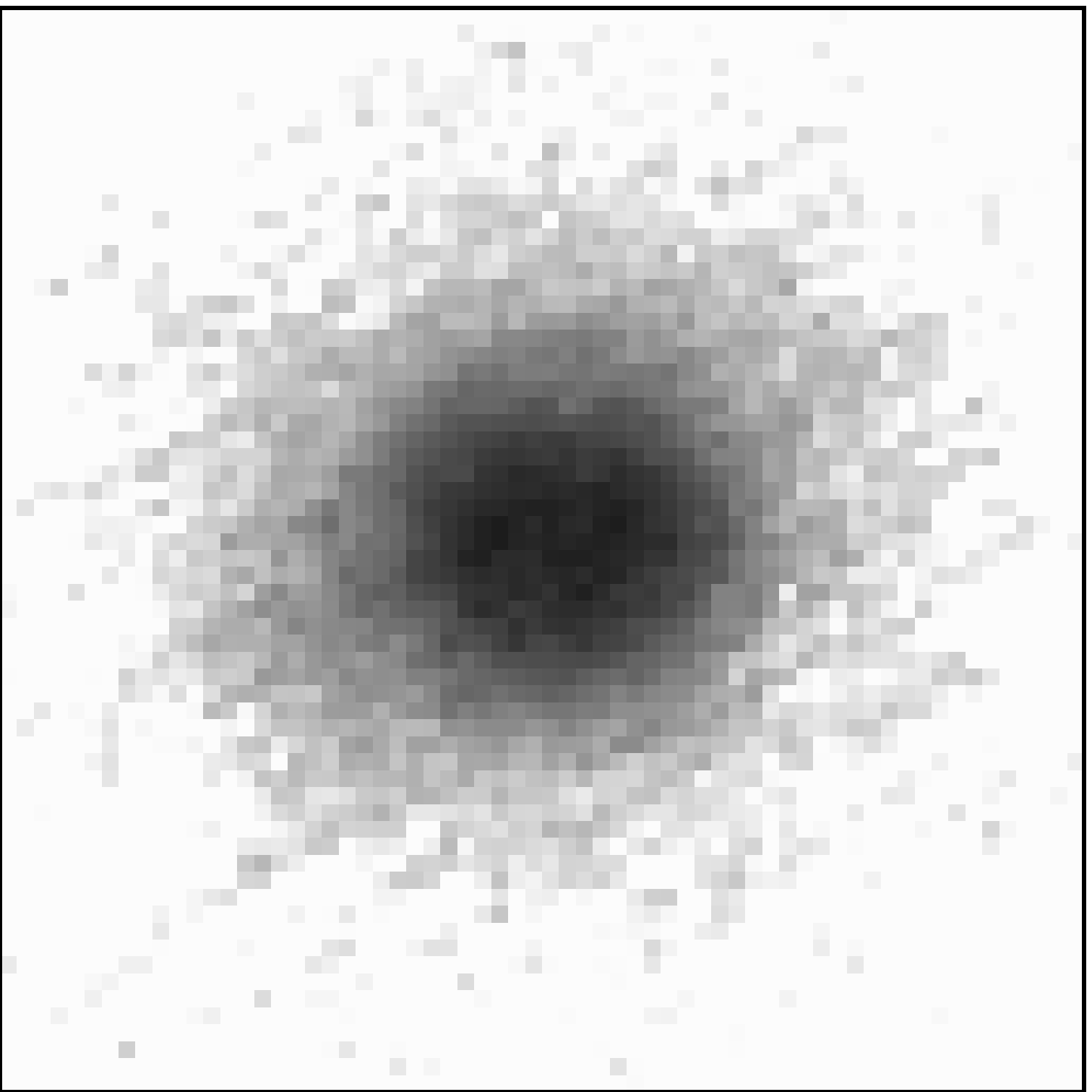}
\epsscale{0.25}
\plotone{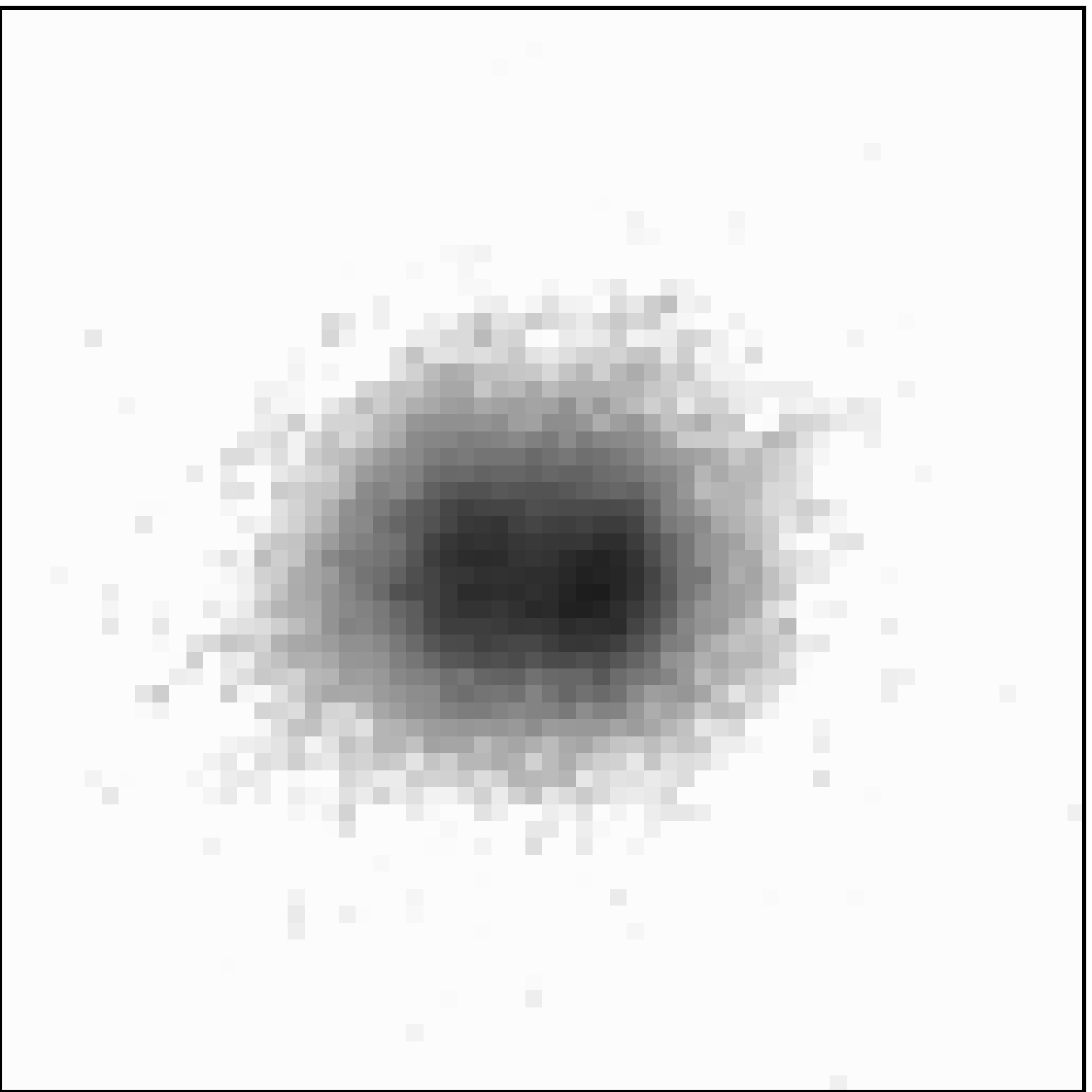}
\epsscale{0.25}
\plotone{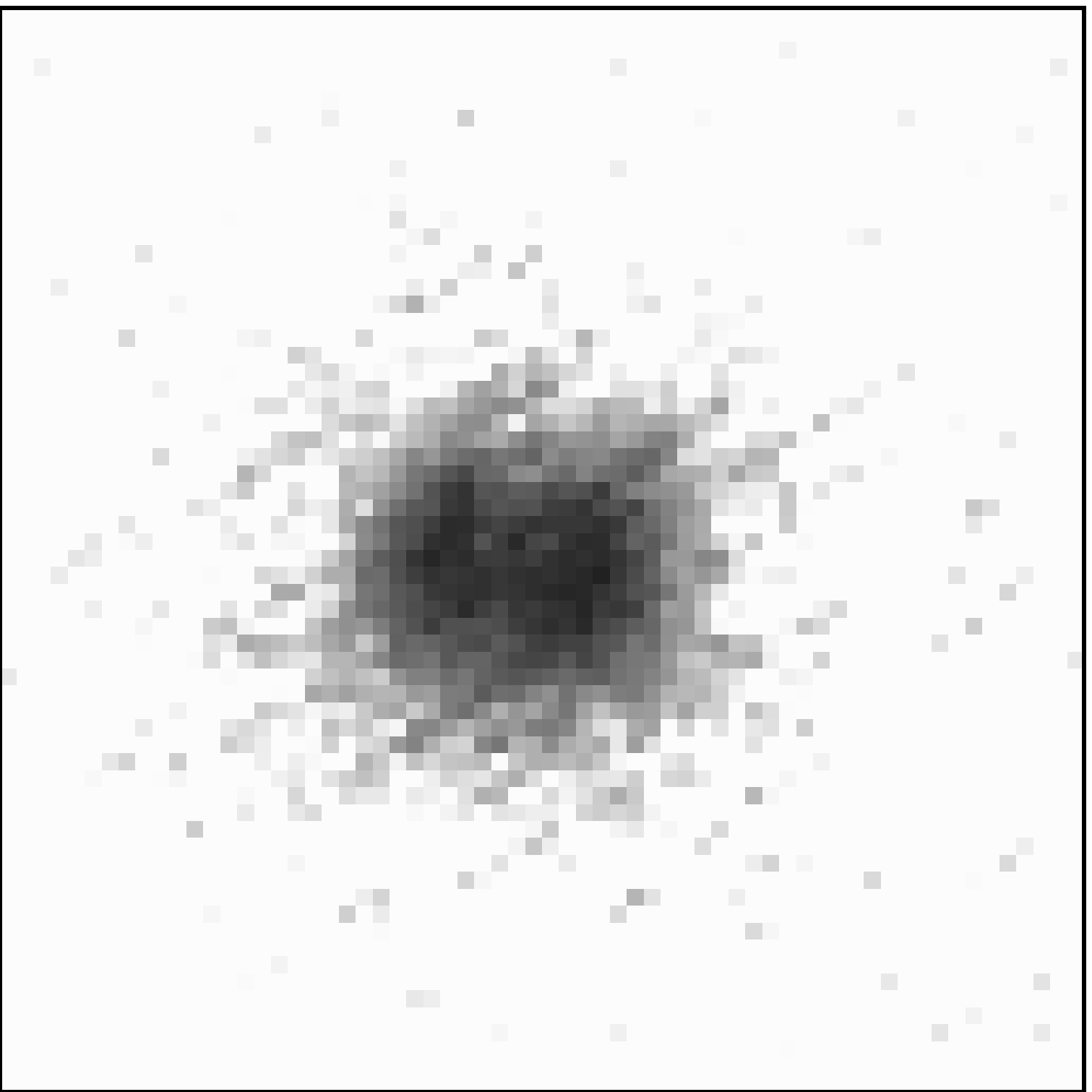}
\caption{$JHK_s$ LGS-AO images (left to right) of WISE 1841+7000AB.  Each image is $\approx$0$\farcs$6 on a side with North up and East to the left.\label{fig_1841mosaic}}
\end{figure}

\begin{figure}
\epsscale{1.0}
\plottwo{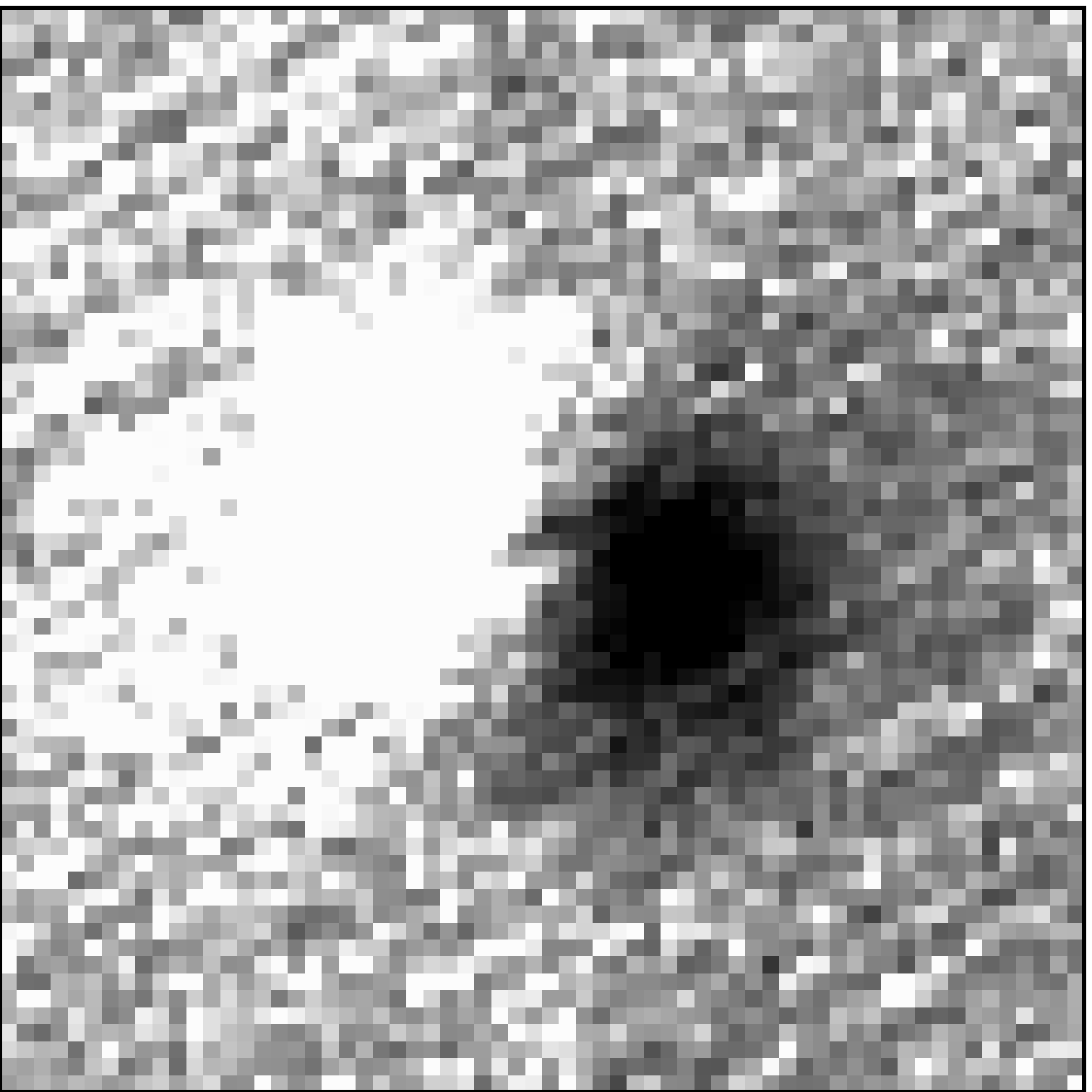}{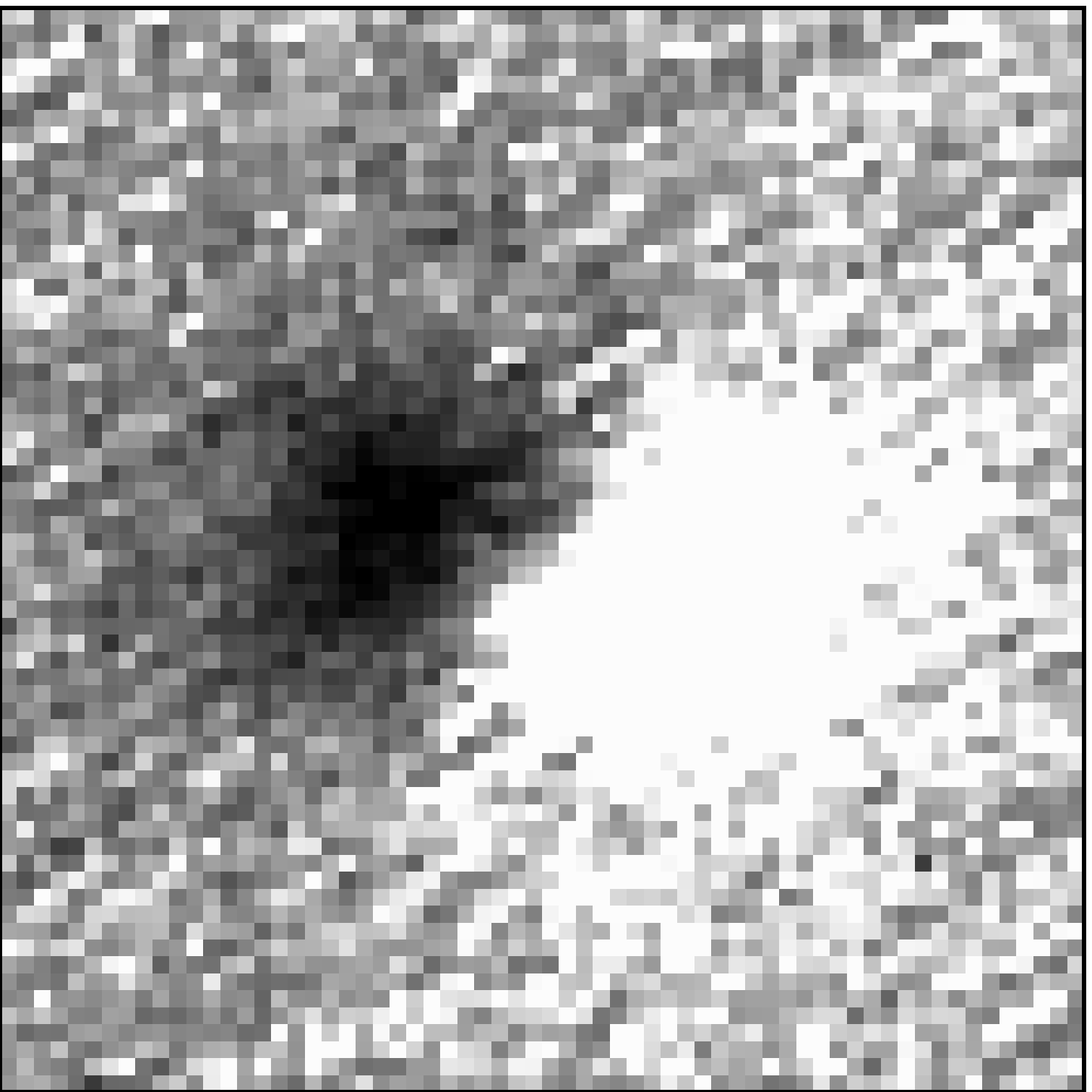}
\caption{Companion subtracted $J$ images for WISE 1841+7000AB. The left panel shows the resultant image after rotating the image about the centroid of the fainter companion and subtracting.  The right panel shows the image after rotating and subtracting the brighter component.  Both images are displayed at the same scale for comparison.  The rotation axis is midway between the positive (shown dark) and negative (shown bright) versions of the object whose photometry is to be measured.
\label{fig_1841_rotsub}}
\end{figure}

\clearpage

\begin{figure}
\epsscale{1.0}
\plottwo{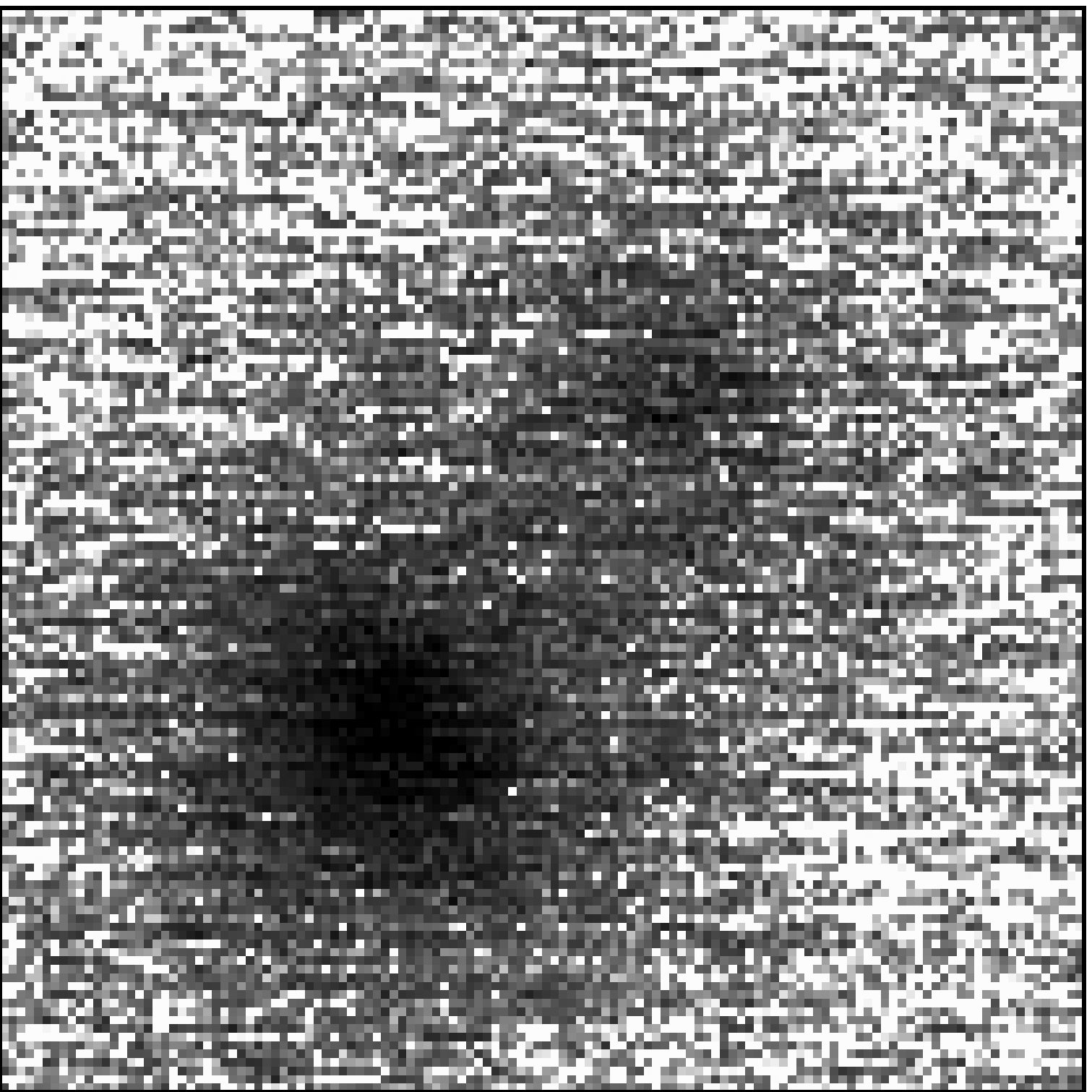}{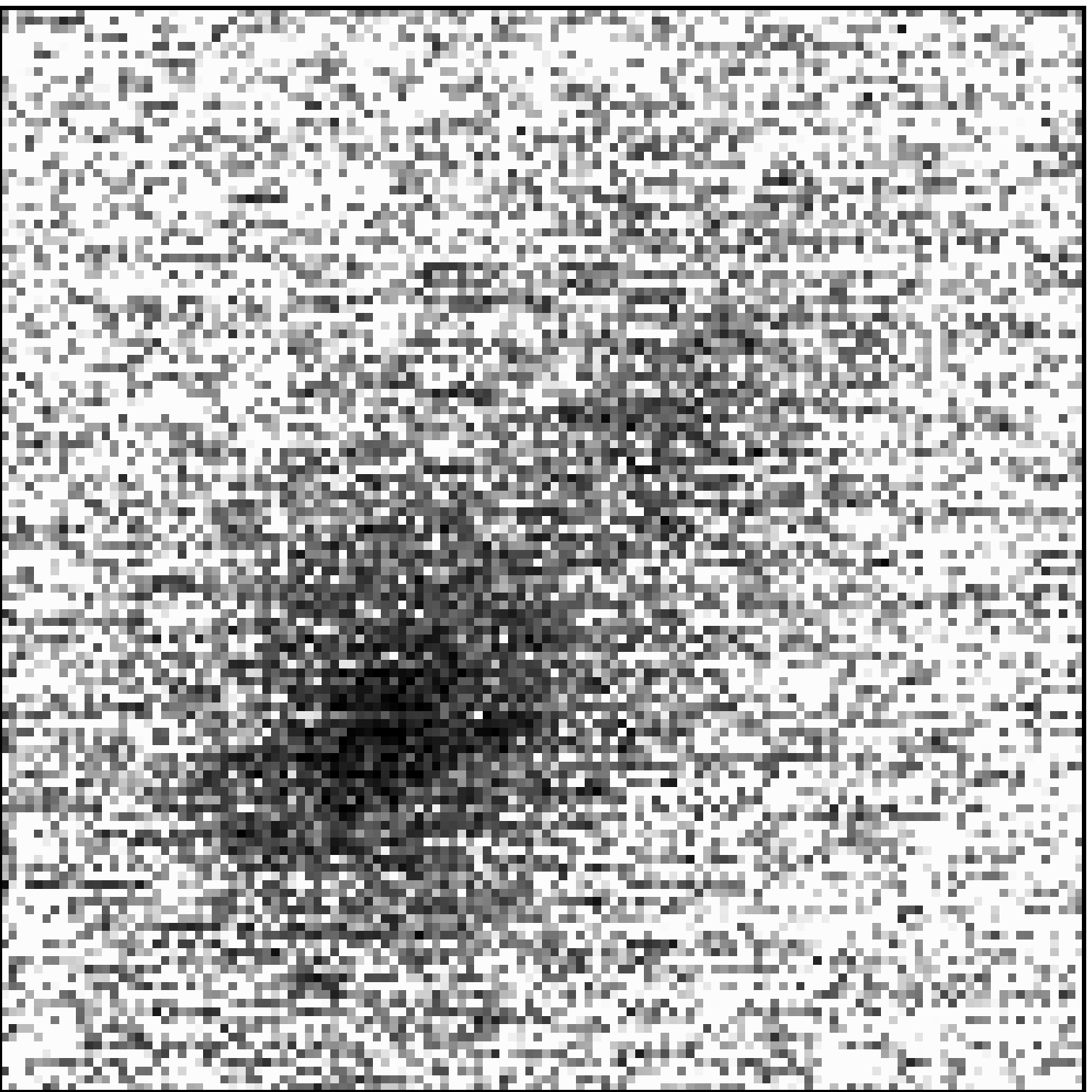}
\caption{$J$ (left) and $H$ (right) LGS-AO images of WISE 0458+6434AB.  The images are $\approx$1.25$\arcsec$ on a side with North up and East to the left.\label{fig_0458mosaic}}
\end{figure}

\begin{figure}
\epsscale{1.0}
\plotone{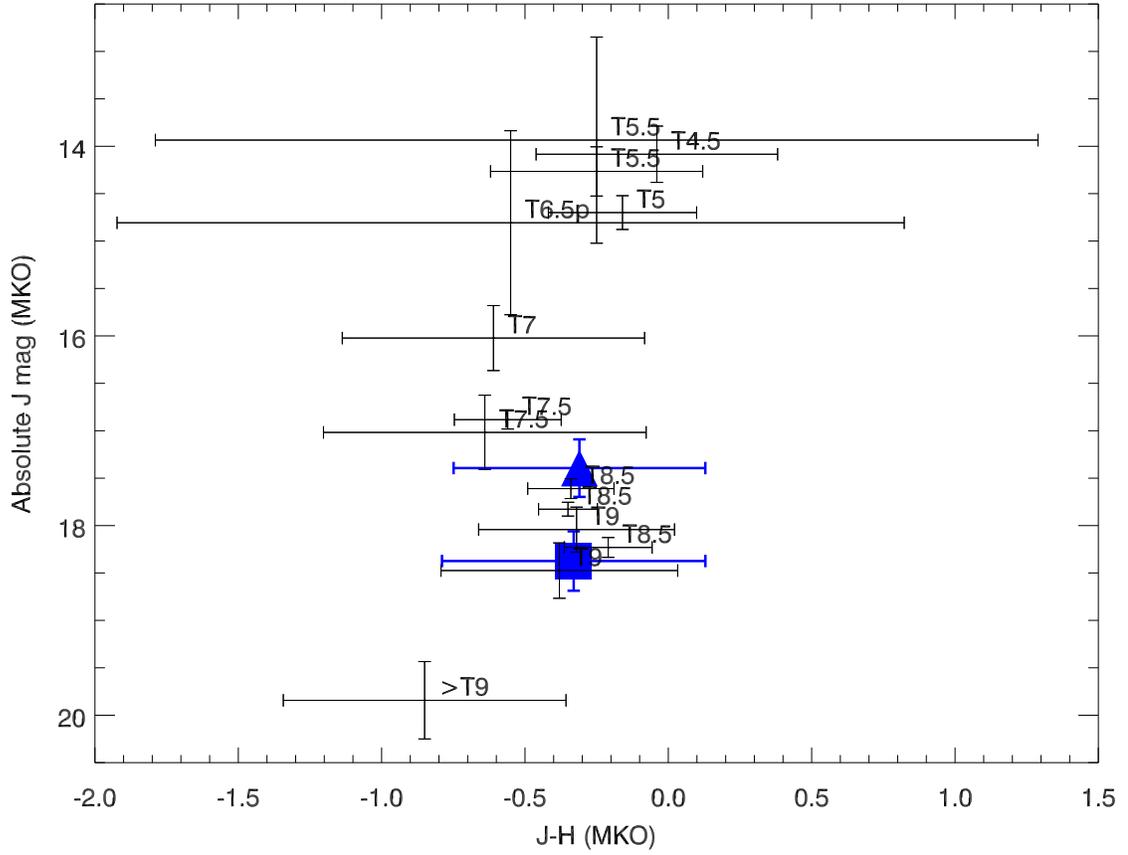}
\caption{Absolute $J$ magnitude as a function of $J-H$ color for WISE 0458+6434 A (blue triangle) and B (blue square) and late-type T dwarfs with measured parallaxes \citep{lucas2010,marocco2010,liu2011}.  The $J-H$ colors and estimated absolute magnitudes of WISE 0458+6434AB are consistent with objects of spectral type $\approx$T8.5--T9, confirming that these are very cold brown dwarfs.\label{fig_0458tcolors}}
\end{figure}


\begin{deluxetable}{lcccccc}
\tabletypesize{\footnotesize}
\tablecaption{Targets\label{tab_targs}}
\tablewidth{0pt}
\tablehead{
  \colhead{Name} &
  \colhead{Spectral} &
  \colhead{W1$-$W2} &
  \colhead{$J$} &
  \colhead{$H$} &
  \colhead{Reference\tablenotemark{a}} &
  \colhead{Near-IR Filter} \\
  \colhead{} &
  \colhead{Type} &
  \colhead{(mag)} &
  \colhead{(mag)} &
  \colhead{(mag)} &
  \colhead{} &
  \colhead{System} \\
}
\startdata
WISEPA J045853.90+643452.6    & T8.5 & 3.38$\pm$0.11& 17.47$\pm$0.07& 17.41$\pm$0.11 & 1,2,3 & 2MASS \\
WISEPA J075003.78+272544.8    & T9   & $>$3.89            & 18.69$\pm$0.04  & 19.00$\pm$0.06 & 2 & MKO \\
WISEPA J132233.67$-$234017.0 & T8 & 3.24$\pm$0.21 & 17.21$\pm$0.10 & 17.01$\pm$0.14 & 2 & 2MASS \\
WISEPA J161441.46+173935.3    & T9 & 4.07$\pm$0.46 & 19.08$\pm$0.06 & 18.47$\pm$0.22 & 2 & MKO \\
WISEPA J161705.75+180714.0    & T8 & 2.86$\pm$0.17 & 17.66$\pm$0.08& 18.23$\pm$0.08 & 2 & MKO \\
WISEPA J162725.64+325524.1    & T6 & 2.76$\pm$0.10 & 16.48$\pm$0.04 & 16.40$\pm$0.05 & 2 & 2MASS \\
WISEPA J165311.05+444423.0    & T8 & 2.81$\pm$0.09 & 17.59$\pm$0.03 & 17.53$\pm$0.05 & 2 & 2MASS \\
WISEPA J174124.27+255319.6    & T9 & 2.94$\pm$0.05 & 16.48$\pm$0.02 & 16.24$\pm$0.04 & 2 & 2MASS \\
WISEPA J184124.73+700038.0    & T5 & 2.26$\pm$0.09 &16.64$\pm$0.03 &16.99$\pm$0.04 & 2 & MKO \\
\enddata
\tablenotetext{a}{(1) \citet{mainzer2011}; (2) Kirkpatrick et al., submitted; (3) Cushing et al. (in prep.)}
\end{deluxetable}

\begin{deluxetable}{cccccc}
\tabletypesize{\footnotesize}
\tablecaption{Observation Log\label{tab_obs}}
\tablewidth{0pt}
\tablehead{
  \colhead{Name} & 
  \colhead{Date (UT)} & 
  \colhead{Reference Star} & 
  \colhead{Filter} & 
  \colhead{Integration (sec)} &
  \colhead{Airmass} \\
}
\startdata
WISE 0458+6434     & 2010 Mar 24 & 1545-0122611 & $H$ & 720 & 1.74-1.80\\
  &  &  &  $J$  & 720 & 1.8-2.0 \\
WISE 0750+2725     & 2010 Dec 27 & 1174-0192590  & $H$ & 720  & 1.01 \\
  &  &  &  $J$  &  720 & 1.00 \\
WISE 1322$-$2340 & 2010 Jul 1 & 0663-0283988 & $H$ & 720 & 1.54 \\
WISE 1614+1739  & 2010 Jul 1 & 1076-0314736 & $H$ & 720 & 1.00 \\
   &   & & $J$ & 720 & 1.01 \\
WISE 1617+1807 & 2010 Jul 1 & 1081-0284597 & $H$ & 360 & 1.03 \\ 
WISE 1627+3255 & 2010 Mar 24 & 1229-0311682 & $H$ & 720 & 1.02 \\
WISE 1653+4444 & 2010 Jul 1 & 1347-0281393 & $H$ & 360 & 1.11 \\
WISE 1741+2553 & 2010 Mar 24 & 1158-0263562 & $H$ & 240 & 1.00 \\
WISE 1841+7000 & 2010 Jul 1 & 1600-0124807 & $H$ & 360 & 1.61 \\
  &  &  &  $J$ & 360 & 1.62 \\
  &  &  &  $K_s$ & 720 & 1.63 \\
\enddata
\end{deluxetable}

\begin{deluxetable}{ccc}
\tablewidth{0pt}
\tablecaption{Properties of WISE 1841+7000AB and WISE 0458+6434AB\label{tab_binprop}}
\tablehead{
  \colhead{Parameter} &
  \colhead{WISE 1841+7000AB} &
  \colhead{WISE 0458+6434AB} \\
}
\startdata
MKO $\Delta J$ (mag) & 0.33$\pm$0.17  &  0.98$\pm$0.08 \\
MKO $\Delta H$ (mag) & 0.02$\pm$0.12  & 1.00$\pm$0.09  \\
MKO $\Delta K_s$ (mag) & 0.10$\pm$0.09 & \nodata \\
MKO $J_{\rm A}$ (mag) & 17.24$\pm$0.10 & 17.50$\pm$0.09  \\
MKO $J_{\rm B}$ (mag) & 17.57$\pm$0.13  & 18.48$\pm$0.12 \\
MKO $H_{\rm A}$ (mag) & 17.73$\pm$0.10 & 17.81$\pm$0.13 \\
MKO $H_{\rm B}$ (mag) & 17.75$\pm$0.10 & 18.81$\pm$0.17 \\
Est. Distance (pc)   & 40.2$\pm$4.9  &   10.5$\pm$1.4\\
$\rho$ (mas) & 70$\pm$14  &  510$\pm$20 \\
$\rho$ (AU) & 2.8$\pm$0.7  &  5$\pm$0.4 \\
$\theta$ ($\degr$) & 82$\pm$9   &  320$\pm$1\\
Est. Orbit Period (yr)\tablenotemark{a} & $\sim$11  & $\sim$70 \\
\enddata
\tablenotetext{a}{For an assumed system age of 1 Gyr.}
\end{deluxetable}

\begin{deluxetable}{lcccc}
\tablewidth{0pt}
\tablecaption{Objects Used to Compute 2MASS-MKO Magnitude Offsets\label{tab_offsets}}
\tablehead{
  \colhead{Name} &
  \colhead{Spectral} &
  \colhead{$J_{2MASS}-J_{MKO}$} &
  \colhead{$H_{2MASS}-H_{MKO}$} &
  \colhead{Discovery/SpT} \\
  \colhead{} &
  \colhead{Type} &
  \colhead{(mag)} &
  \colhead{(mag)} &
  \colhead{Reference\tablenotemark{a}} \\
}
\startdata
2MASS J04151954$-$0935066   &    T8  & +0.329   & $-$0.026 & 1/2 \\
2MASS J09393548$-$2448279   &    T8  &  +0.329  & $-$0.044  & 3/2 \\
Wolf 940B                                          & T8.5  &  +0.339  & $-$0.033  & 4/5 \\
ULAS J003402.77$-$005206.7    & T8.5  & +0.352   & $-$0.038  &  6/5 \\
ULAS J133553.45+113005.2        & T8.5  &  +0.346  &  $-$0.047 &  7/5 \\
CFBDS J005910.90$-$011401.3 & T8.5  & +0.340   &  $-$0.043  &  8/5 \\
UGPS 072227.51$-$054031.2     &  T9     &  +0.335  &  $-$0.042  & 9/5  \\
\enddata
\tablenotetext{a}{(1) \citet{burgasser2002}; (2) \citet{2006ApJ...637.1067B}; (3) \citet{2005AJ....130.2326T}; (4) \citet{burningham2009}; (5) Cushing et al. (in prep.);  
 (6) \citet{warren2007}; (7) \citet{burningham2008}; (8) \citet{delorme2008}; (9) \citet{lucas2010}}
\end{deluxetable}

\clearpage

\bibliographystyle{apj}
\bibliography{ms}

\end{document}